\documentclass[pra,twocolumn,showpacs,amsmath,amssymb]
{revtex4}

\usepackage[dvips]{color}
\usepackage{graphicx}

\begin{document}

\title{Isentropes of spin-1 bosons in an optical lattice}

\author{K. W. Mahmud$^1$, G. G. Batrouni$^2$, R. T. Scalettar$^1$}
\affiliation{$^1$Physics Department, University of California, Davis,
California 95616, USA \\
$^2$INLN, Universit\'e de Nice--Sophia Antipolis, CNRS; 1361 route des Lucioles, 06560 Valbonne, France}

\begin{abstract}
We analyze the effects of adiabatic ramping of optical lattices on the
temperature of spin-1 bosons in a homogeneous lattice. Using
mean-field theory, we present the isentropes in the
temperature-interaction strength ($T,U_0$) plane for ferromagnetic,
antiferromagnetic, and zero spin couplings. Following the isentropic
lines, temperature changes can be determined during adiabatic loading of
current experiments. We show that the heating-cooling separatrix lies on
the superfluid-Mott phase boundary with cooling occuring within the
superfluid and heating in the Mott insulator, and quantify
the effects of spin coupling on the heating rate. We find that the
mean-field isentropes for low initial entropy terminate at the
superfluid-Mott insulator phase boundary.
\end{abstract}


\maketitle

%%%%%%%%%

\section{Introduction}

Ultra-cold atoms in optical lattices offer the possibility of
realizing various fundamental models of strongly correlated bosons and
fermions~\cite{greiner02,leweinstein07}. A crucial aspect of the
optical lattice system is its flexibility in controlling different
lattice parameters and particle interactions, thereby facilitating
progress towards the creation of quantum emulators. After the
observation of the superfluid to Mott insulator transition with spin-0
bosons~\cite{greiner02}, steady progress has been made towards
trapping spinful atoms. These models, along with a precise knowledge
and tunability of the microscopic Hamiltonian can potentially lead to
a better understanding of quantum magnetism, and related phenomena.
Unlike magnetic traps which freeze the $F_z$ component of spin,
optical traps can confine $^{23}$Na, $^{39}$K, and $^{87}$Rb with
hyperfine spin $F=1$. Several theoretical studies have focused on
spinor condensates in an optical
lattice~\cite{demler02,snoek04,ho98,batrouni09,pai08} and how the spin
degree of freedom modifies the phase diagram and the nature of
superfluid-Mott insulator transition.  Experiments have also explored
the properties of spinful bosons in harmonic traps~\cite{kurn98} and,
recently, in a double well optical superlattice~\cite{trotzky08}.

The technological breakthrough in cooling to ultra-cold temperatures
paved the way for the realization of Bose-Einstein condensation and
optical lattice experiments. The temperature of a bosonic gas
in a trap can be measured accurately. However, no established temperature
measurement exists for optical lattice systems, although several proposals
have been made~\cite{ho0908,zhou09}. This makes it
difficult to obtain a quantitative description of the various low
temperature phases and thermal and quantum phase transitions between
them~\cite{diener07}.

In current experiments, ultra-cold atoms are first loaded in a
harmonic trap, and then an external sinusoidal potential created by
interfering lasers are slowly ramped up to create the optical
lattice. For gradual enough ramping, there is no heat exchange with
the environment, and this process can be considered adiabatic
(constant entropy)~\cite{greiner02,spielman07}. It is of great interest to the
experimentalists to know how the system cools down or heats up during
the adiabatic process. The change in temperature with adiabatic
ramp-up of optical lattice for the spinless Bose-Hubbard model and the
Fermi Hubbard model has been studied by several
authors~\cite{rey06,yoshimura08,pollet08, werner05,paiva09}.

In this paper, we present the isentropes for spin-1 bosons in a
homogeneous optical lattice for both ferromagnetic and antiferromagnetic
spin couplings. We investigate the effects of adiabatic ramping on
temperature within the mean-field approximation, and show that for
spinor bosons cooling occurs in the superfluid phase and heating
occurs in the Mott insulator or normal phase. We find that the
heating-cooling separatrix lies along the superfluid-Mott phase
boundary. As the magnitude of spin coupling increases, the rate of
temperature change decreases in the Mott regime, and can both increase
and decrease in the superfluid regime depending on the value of the
spin coupling. We find that the mean-field isentropes for low initial
entropy terminate at the superfluid-Mott insulator phase boundary and
argue that this is a consequence of the absence of breaking of the
degeneracy of the ground state in the mean field approximation.

The paper is organized as follows. We discuss the spin-1 Bose-Hubbard
model and the details of the mean-field theory in section II. In
section III, we investigate the entropy for this model, and present
our results for the isentropes and temperature changes for different
couplings.  A summary is contained in section IV.

%%%%%%%%%%%%%%  FIGURE  %%%%%%%%%%%%%%%%%%%%%%%%%%%%%%%%
\begin{figure}[h]
\begin{center}
  \includegraphics[width=0.35\textwidth,angle=0]{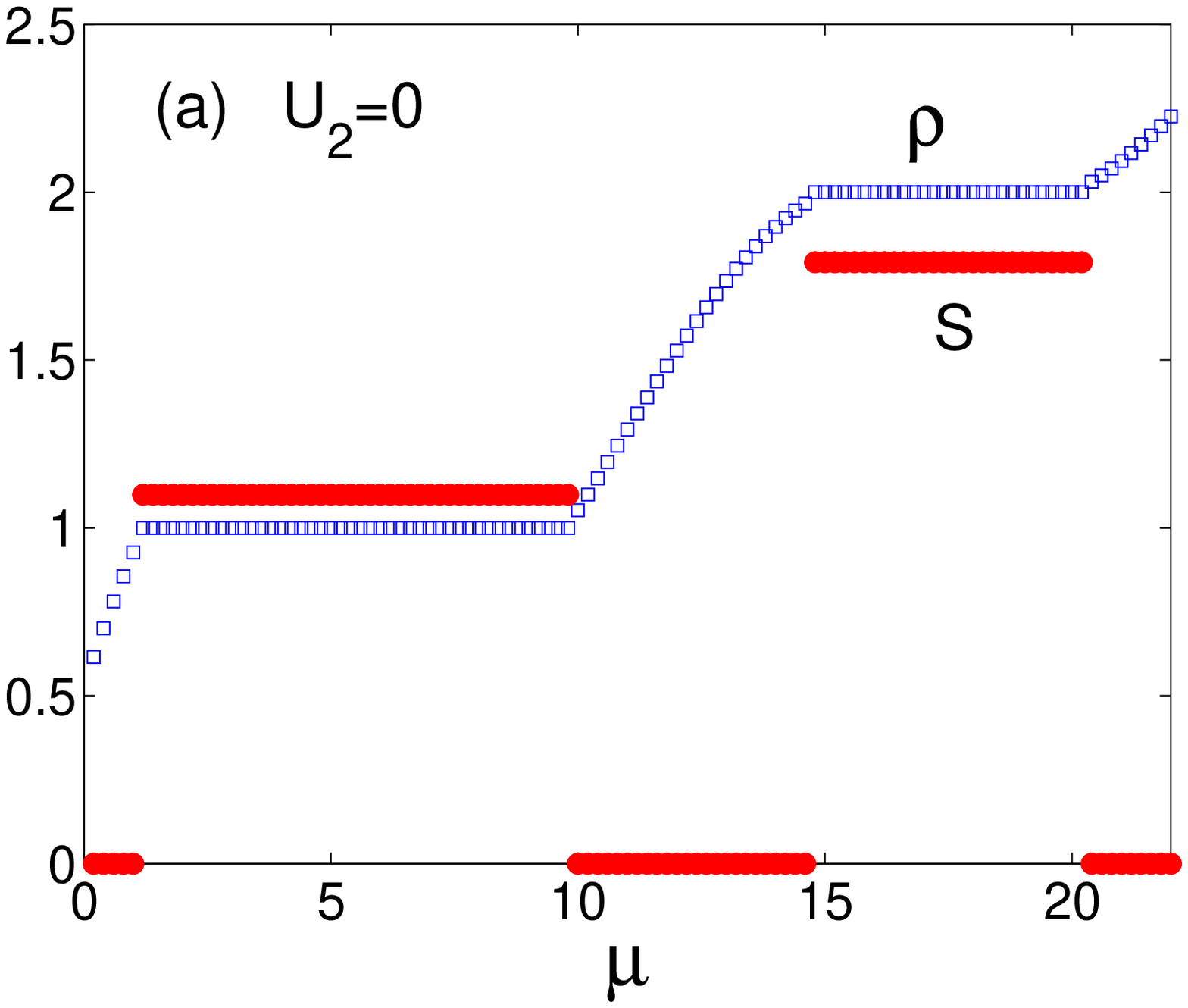}
\end{center}
\begin{center}
  \includegraphics[width=0.35\textwidth,angle=0]{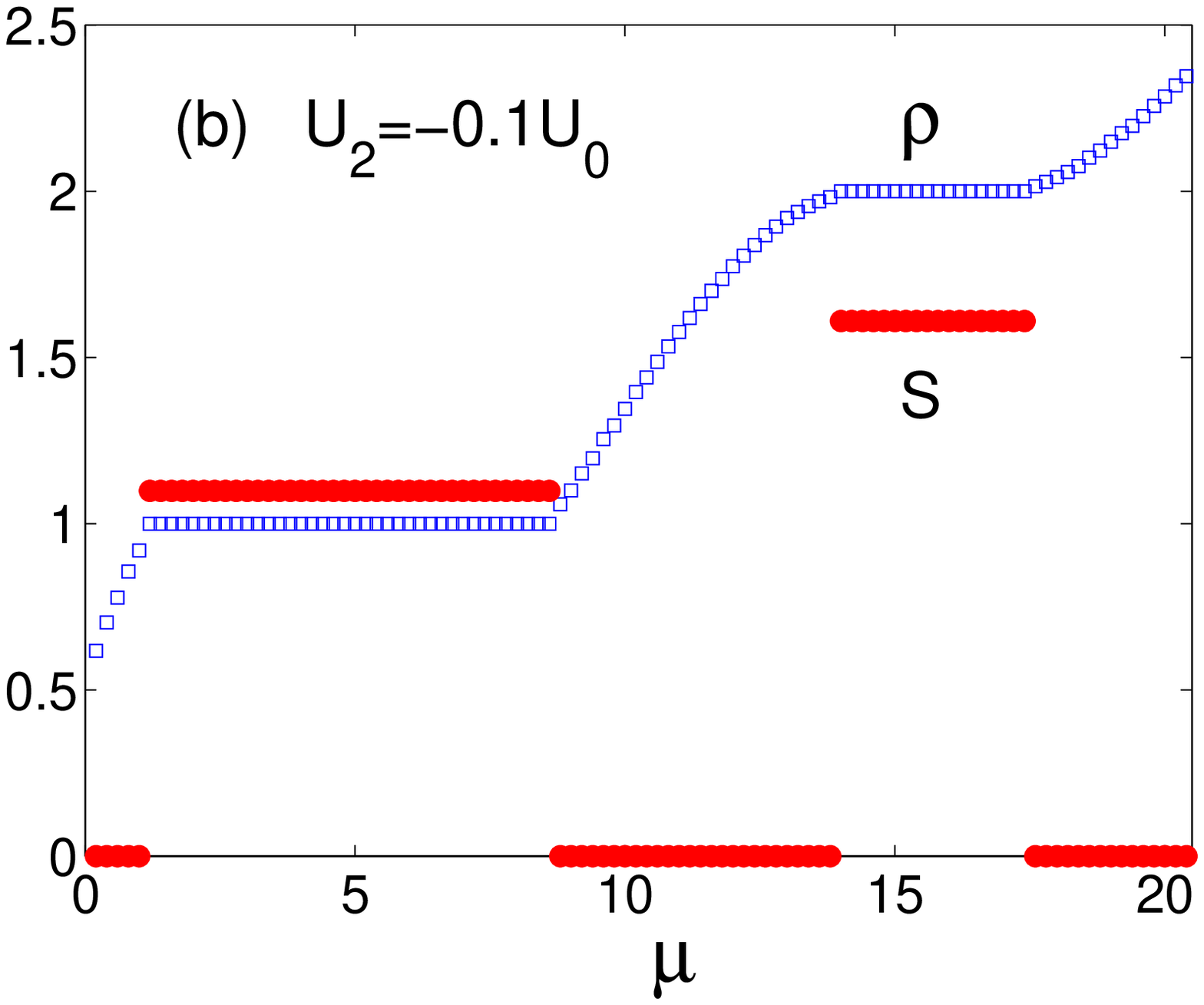}
\end{center}
\begin{center}
  \includegraphics[width=0.35\textwidth,angle=0]{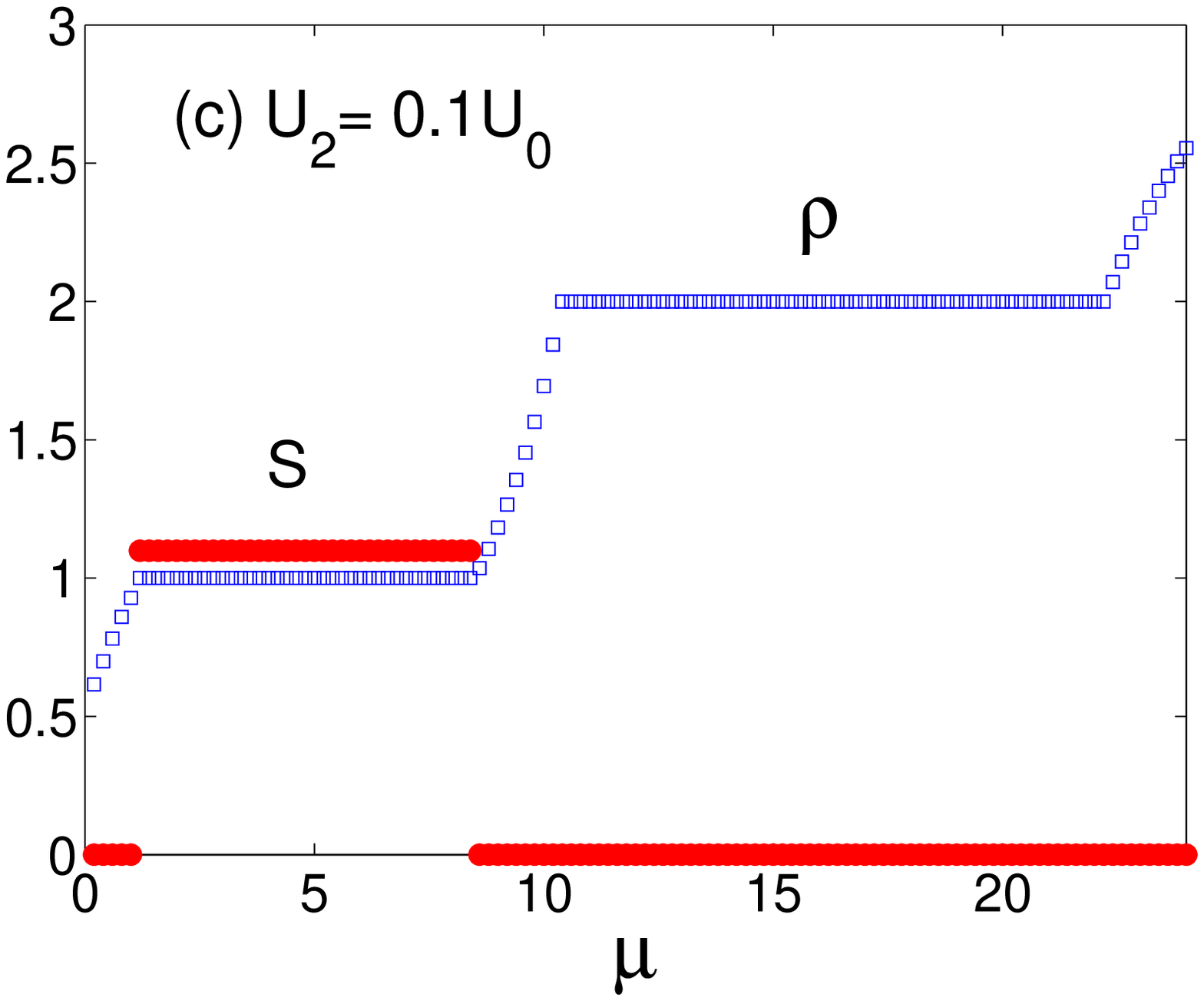}
\end{center}
\vspace{-0.6cm}
\caption{\label{fig:entropyU2} (color online).  Entropy and filling as
functions of chemical potential at zero temperature for $U_0=12 zt$ (we set
$zt=1$ in our calculation) and
(a) $U_2=0$, (b) $U_2=-0.1U_0$ and (c) $U_2=0.1U_0$. For $U_2=0$, the
entropy is $\ln(3)$ at $\rho=1$, $\ln(6)$ at $\rho=2$, and zero
elsewhere. This can be understood in terms of the degeneracy of spin-1
(three component) bosons and total spin at each site. For $U_2<0$, the
energy is minimized with maximum total spin $F=2$, and hence the
entropy at $\rho=2$ changes to $\ln(5)$. Similarly, for $U_2>0$, the
ground state has lowest total spin, $F=0$, and hence the system has
zero entropy for $\rho=2$.  }
\end{figure}
%%%%%%%%%%%%%%%%%%%%%%%%%%%%%%%%%%%%%%%%%%%%%%%%%%%%%%%%

\section{Spin-1 Bose-Hubbard Model and Mean-Field theory}

The Hamiltonian for spin-1 bosons in an optical lattice is given by
\begin{eqnarray}
H &=& -t \sum_{\langle i,j \rangle,\sigma} \left(
a^\dagger_{i\sigma} a^{}_{j\sigma} + a^\dagger_{j\sigma} a^{}_{i\sigma} \right)
+ \dfrac{U_o}{2} \sum_i \hat{n}_{i} \left( \hat{n}_{i} -1 \right) \nonumber \\
&& + \dfrac{U_2}{2} \sum_i \left( \vec{F}^{2}_{i}
-2\hat{n}_{i}\right) - \sum_i \mu_{i} \hat{n}_{i},
\label{hamil}
\end{eqnarray}
Here $a^\dagger_{i\sigma}$ ($a_{i\sigma}$) are boson creation
(destruction) operators at site $i$ with spin component $\sigma$
($\sigma = 1, 0, -1$). The first term in Eq.(\ref{hamil}) describes
the spin dependent hopping between near neighboring sites. In the
second term, $U_o$ is the on-site repulsion, and
$\hat{n}_{i}=\sum_{\sigma} a^\dagger_{i\sigma} a_{i\sigma}$ counts the
total number of bosons on site $i$. In the third term, $U_2$ is the
spin dependent interaction which can be zero, positive or negative,
and $\vec{F}_{i}=\sum_{\sigma,\sigma'}a^\dagger_{i\sigma}
\vec{F}_{\sigma \sigma'} a_{i\sigma'}$ is the total spin on site $i$,
where $\vec{F}_{\sigma \sigma'}$ are the standard spin-1 matrices.

The spin dependent interaction, $U_2$, for the spin-1 model greatly
modifies its physics compared to that of spin-0 bosons~\cite{pai08}.
From the symmetry of the bosonic wavefunction, scattering with total
spin $F=1$ is prohibited. The difference in scattering lengths in the
$F=0$ ($a_0$) and $F=2$ ($a_2$) channels is responsible for the spin
dependent coupling. The interactions can be expressed as $U_0=4\pi
\hbar^2 (a_0+2a_2)/3M$ and $U_2=4\pi \hbar^2 (a_2-a_0)/3 M$, $M$ being
the mass of the atom. The spin dependent interaction is ferromagnetic
when $U_2<0$ ($a_2<a_0$) and antiferromagnetic when $U_2>0$
($a_2>a_0$).  $^{23}$Na atoms are ferromagnetic and $^{87}$Rb
antiferromagnetic. Our study in this paper considers both signs of the
interaction.  The coupling constants obey the constraint
$-1<U_2/U_0<1/2$.

The zero-temperature phase diagram for the spin one model has been
calculated with numerical methods such as QMC
\cite{batrouni09,apaja06}, DMRG \cite{rizzi05}, and also within the
mean-field approximation \cite{pai08,sheshadri93}.  As for the spin-0
case, mean-field theory for the spin-1 Bose-Hubbard
model~\cite{pai08,krutitsky04,kimura05} captures the essential physics
of this system. A finite temperature extension of the mean-field
analysis was presented in \cite{pai08}, revealing the rich phase
diagram that includes both first and second order transitions.  The
spinor Bose-Hubbard model, with filling of one boson per site and for
small hopping, can be mapped onto the $F=1$ bilinear biquadratic
Heisenberg model, which has been studied by many
authors~\cite{yip08,chubukov91,imambekov03} to gain an understanding
of different Mott phases. Here, we will extend the mean-field
calculation presented in Ref.\cite{pai08} to obtain the entropy.

%%%%%%%%%%%%%%  FIGURE  %%%%%%%%%%%%%%%%%%%%%%%%%%%%%%%%
\begin{figure}[h]
\begin{center}
  \includegraphics[width=0.35\textwidth,angle=0]{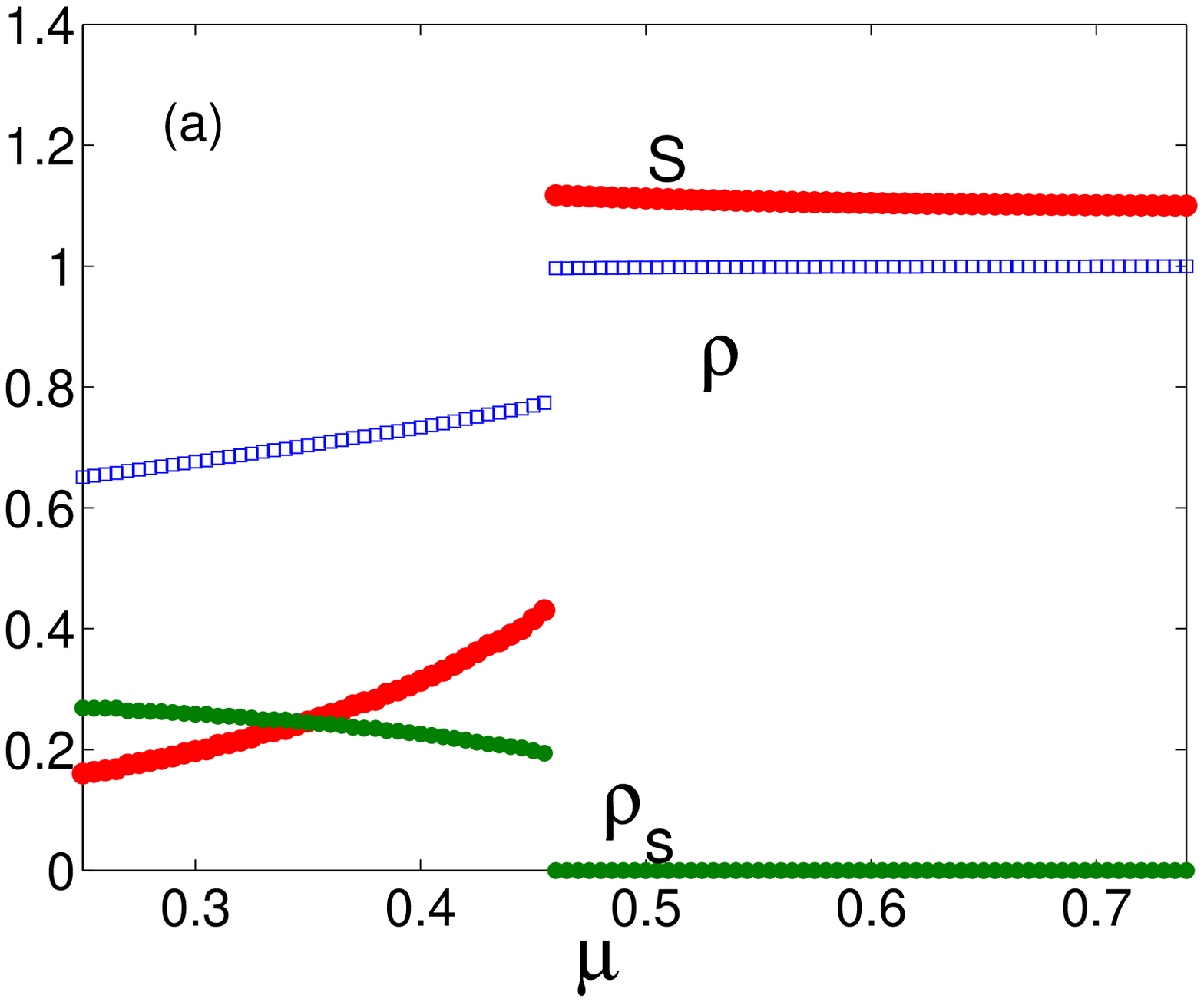}
\end{center}
\begin{center}
  \includegraphics[width=0.36\textwidth,angle=0]{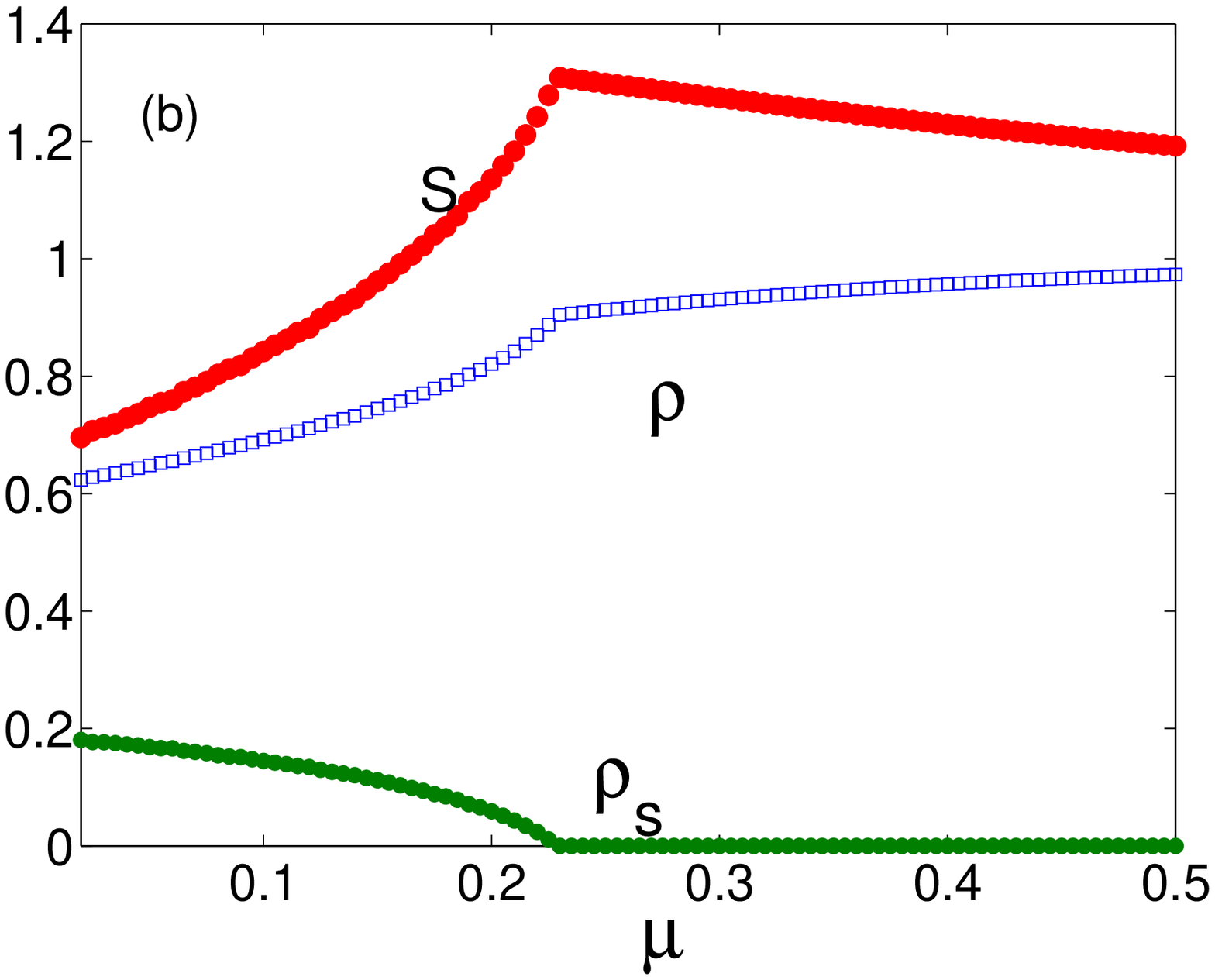}
\end{center}
\vspace{-0.6cm}
\caption{\label{fig:order} (color online).  Traversing a first order
(a) and second order (b) phase boundary in the ($\mu,T$)
plane. $T=0.1$ (a) and $T=0.2$ (b). The discontinuity in the entropy
in (a) satisfies the Clausius-Clapeyron equation relating the entropy
jump to the slope of the first order phase boundary.  Here $U_0=12,
U_2=0$.  }
\end{figure}
%%%%%%%%%%%%%%%%%%%%%%%%%%%%%%%%%%%%%%%%%%%%%%%%%%%%%%%%

In the mean-field approximation, the hopping term is decoupled as
\begin{equation}
a^\dagger_{i\sigma} a_{j\sigma} \simeq \langle
a^\dagger_{i\sigma}\rangle a_{j\sigma}+ a^\dagger_{i\sigma} \langle
a_{j\sigma} \rangle-\langle a^\dagger_{i\sigma}\rangle \langle
a_{j\sigma} \rangle \,\, ,
\label{mfa}
\end{equation}
neglecting fluctuations around the equilibrium value. Here, we define
$\psi_{\sigma}=\langle a^{\dagger}_{j\sigma} \rangle=\langle
a_{j\sigma} \rangle$, for $\sigma=1, 0, -1$ to be the superfluid order
parameter.  The use of Eq.(\ref{mfa}) allows us to rewrite
Eq.(\ref{hamil}) as a sum of independent single site Hamiltonians, $
H=\sum_{i} H^{\rm mf}_{i} $ where
\begin{eqnarray}
H^{\rm mf}_{i} &=& \dfrac{U_o}{2} \hat{n}_{i} \left( \hat{n}_{i} -1
\right) + \dfrac{U_2}{2} \left( \vec{F}^{2}_{i}
-2\hat{n}_{i}\right) \nonumber \\
&& - \mu \hat{n}_{i} - \sum_{\sigma} \psi_{\sigma}
\left(a^\dagger_{i\sigma} + a_{i\sigma} \right) + \sum_{\sigma}
|\psi_{\sigma}|^2. \label{mfham}
\end{eqnarray}
Here we set $zt=1$, where $z$ is the number of nearest neighbors.
To perform the mean field calculations, we write the matrix
elements of the Hamiltonian $H^{\rm mf}_{i}$ in the occupation
number basis $|n_{i,-1},n_{i,0},n_{i,1} \rangle$, and truncate the
onsite Hilbert space $H_i$ by allowing a maximum number of
particles per site, ${\rm N}_{max}=4$, for which the truncation
effects are negligible. We use ${\rm N}_{max}=6$ for
 simulations with higher number density such as in Fig.~\ref{fig:isen3}.
 Since we are treating a homogeneous system, the site index $i$ can
be ignored. We diagonalize the Hamiltonian to obtain the energy
spectrum $E_{\alpha}$ and eigenstates $|\phi_{\alpha} \rangle$,
and evaluate the partition function and the free energy,
\begin{equation}
Z(\mu,U_0,U_2,T;\psi_{\sigma})=\sum_{\alpha} e^{-E_{\alpha}/T}
\end{equation}
and
\begin{equation}
{\cal F}(\mu,U_0,U_2,T;\psi_{\sigma})=-T \ln Z(\mu,U_0,U_2,T;\psi_{\sigma})
\end{equation}
We set $k_B=1$ throughout our calculation. For given $\mu, U_0, U_2$
and $T$, the superfluid order parameters $\psi_{\sigma}$ are obtained
by minimizing the free energy, {\it i.e}.~by solving $\partial {\cal
F}/ \partial \psi_{\sigma}=0$ for $\sigma=1, 0, -1$. Solving for the
extrema of the free energy is equivalent to the self-consistency
condition for $\psi_{\sigma}$.

After determining the values of the superfluid order parameters
$\psi^{\rm eq}_{-1},\psi^{\rm eq}_0$ and $\psi^{\rm eq}_{1}$ that
minimize the free energy, other physical quantities can be obtained
easily from the resulting global minimum of free energy, ${\cal F}^{\rm eq}$,
and eigenvalues ($E^{\rm eq}_\alpha$) and eigenstates ($\phi^{\rm
eq}_\alpha$) at the global minimum.  The superfluid density is given
by,
\begin{equation}
\rho_{S}=\sum_{\sigma}|\psi^{\rm eq}_{\sigma}|^2
\end{equation}
and the number density is,
\begin{equation}
\rho=-\partial {\cal F}/\partial \mu=\frac{1}{Z} \sum_{\alpha}
e^{-E^{\rm eq}_{\alpha}/T}\langle \phi^{\rm
eq}_{\alpha}|\hat{n}|\phi^{\rm eq}_{\alpha} \rangle.
\end{equation}
Finally, the entropy is calculated from
\begin{equation}
S=- \partial {\cal F}/\partial T=\ln Z+\frac{1}{Z T} \sum_{\alpha}
E^{\rm eq}_{\alpha} e^{-E^{\rm eq}_{\alpha}/T}.
\end{equation}
Pai {\it et al}.~\cite{pai08} give a detailed analysis of the
superfluid order parameters $\psi_1, \psi_0, \psi_{-1}$ for different
phases and show that in the antiferromagnetic (polar) superfluid, the
possibilities are $\psi_1=\psi_{-1}>0,\psi_0=0$ and
$\psi_1=\psi_{-1}=0,\psi_0>0$. For the ferromagnetic superfluid,
$\psi_1=\psi_{-1}$, $\psi_0=\sqrt{2}\psi_1$.

We use the superfluid density to determine the phase diagram in the
($T,U_0$) plane. $\rho_s>0$ corresponds to the ferromagnetic or polar
superfluid, and $\rho_s=0$ corresponds to the Mott insulator phase. At
finite temperature, there is a crossover from the Mott phase to the
normal phase depending on the value of the compressibility.

%%%%%%%%%

\section{Entropy and Isentropic curves}

%%%%%%%%%%%%%%  FIGURE  %%%%%%%%%%%%%%%%%%%%%%%%%%%%%%%%
\begin{figure}[ht]
\begin{center}
\includegraphics[width=0.45\textwidth,angle=0]{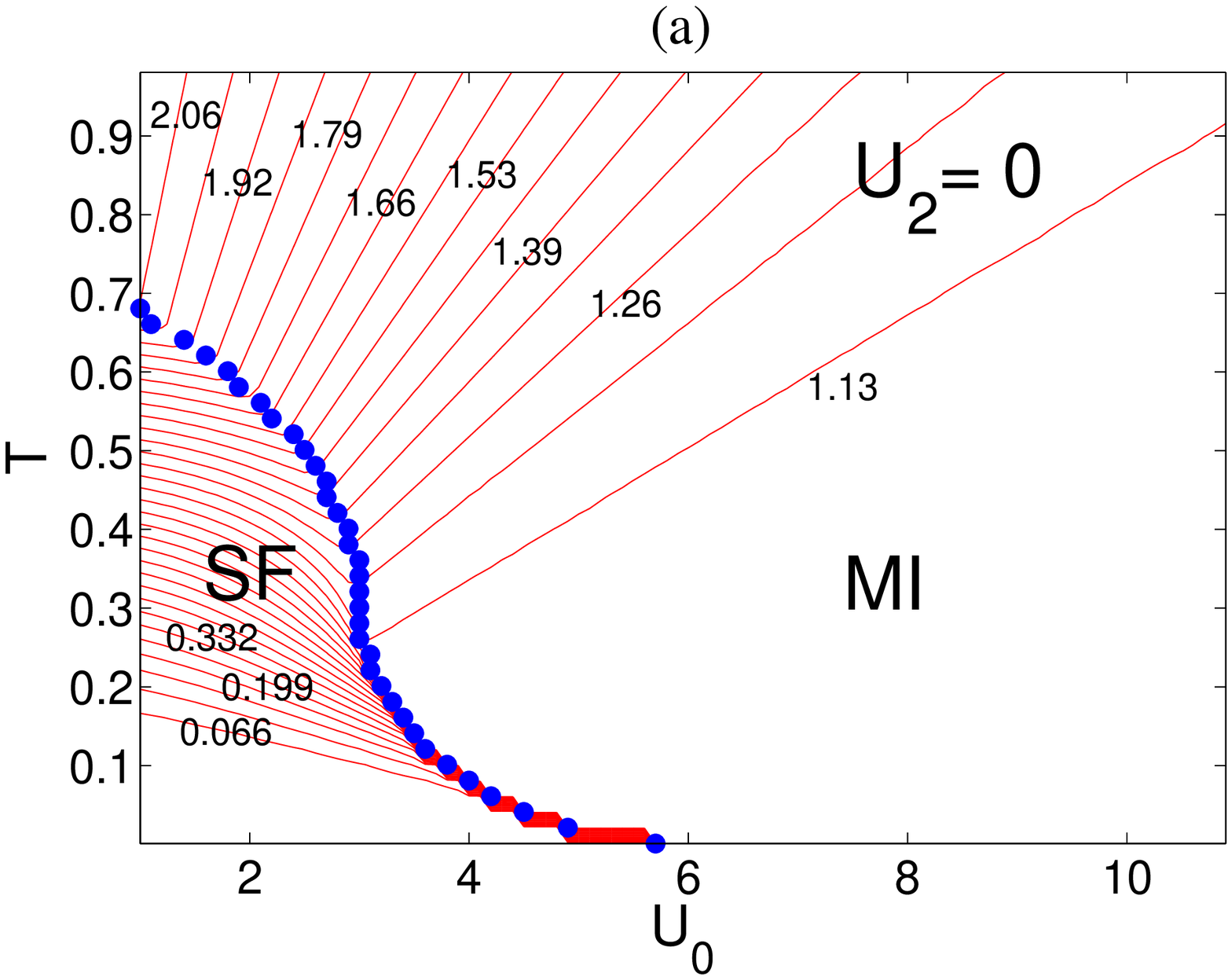}
\end{center}
\begin{center}
  \includegraphics[width=0.4\textwidth,angle=0]{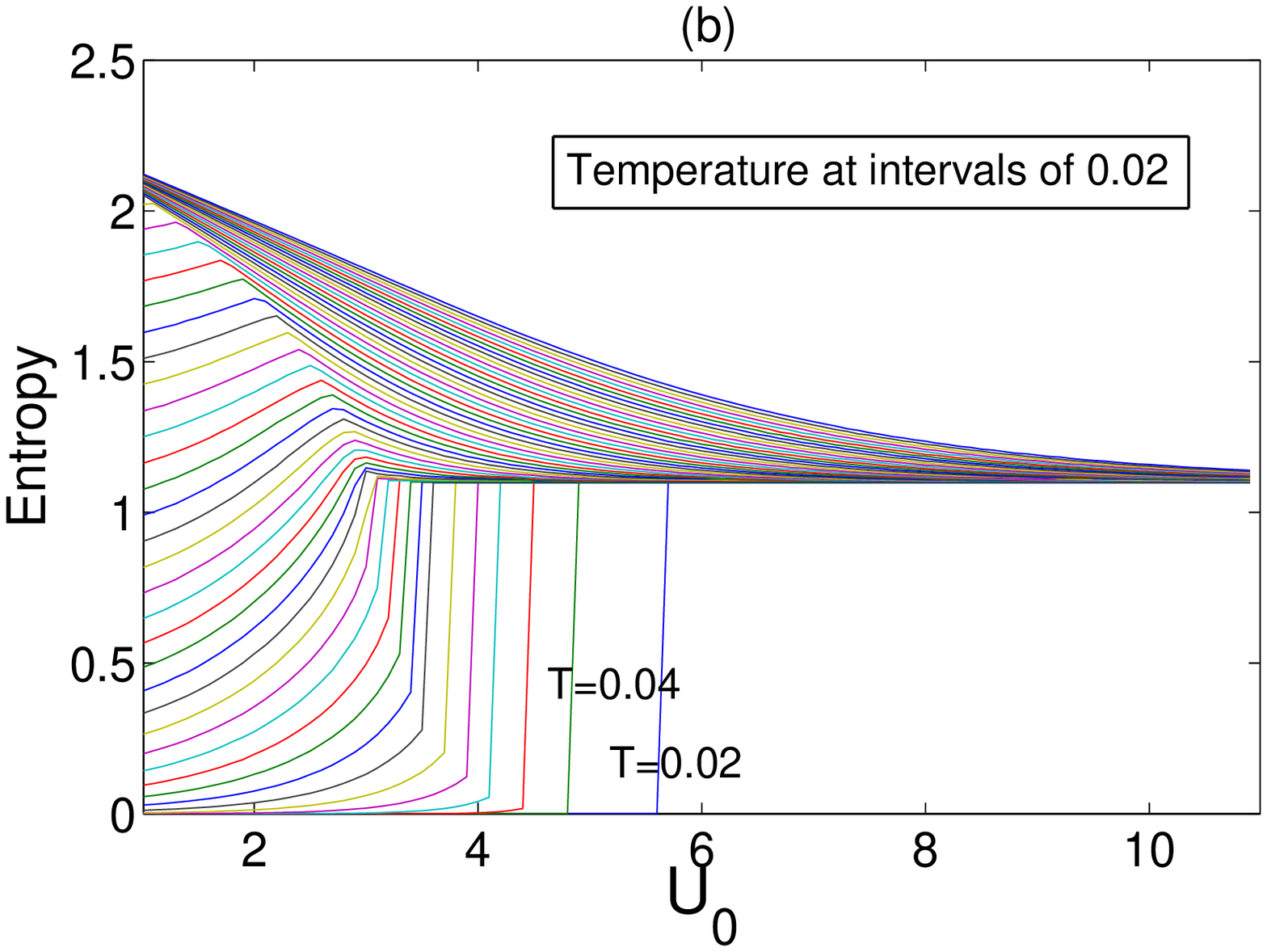}
\end{center}
\vspace{-0.6cm}
\caption{\label{fig:isen1} (color online).  (a) Isentropic curves
overlaid on the finite temperature phase diagram for $U_2=0$ and
density, $\rho=1$. During adiabatic ramping of the lattice, the system
cools along the isentropic lines in the superfluid region, and heats
in the Mott region. In our mean field analysis, the separatrix of
cooling and heating lies on the superfluid-Mott phase boundary in the
($T,U_0$) plane. (b) Shows entropy as a function of $U_0$ at constant
temperatures for intervals of $T=0.02$. The termination of the isentropic
lines at the phase boundary can be understood by the absence of low entropy
states as $U_0$ increases into the Mott regime.}
\end{figure}
%%%%%%%%%%%%%%%%%%%%%%%%%%%%%%%%%%%%%%%%%%%%%%%%%%%%%%%%

%%%%%%%%%%%%%%  FIGURE  %%%%%%%%%%%%%%%%%%%%%%%%%%%%%%%%
\begin{figure}[ht]
\includegraphics[width=0.6\textwidth,angle=0]{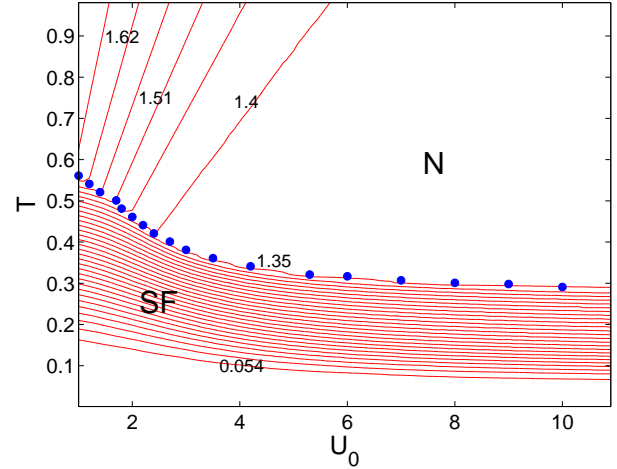}
\caption{\label{fig:noninteger} (color online). Isentropic curves for
a noninteger filling, $\rho=0.7$ and $U_2=0$. As in the commensurate
filling case, the heating-cooling separatrix lies on the phase
boundary. However, the low temperature isentropes do not terminate,
and remain under the phase boundary.}
\end{figure}
%%%%%%%%%%%%%%%%%%%%%%%%%%%%%%%%%%%%%%%%%%%%%%%%%%%%%%%%

Figure~\ref{fig:entropyU2}(a) shows the entropy ($S$) and density
($\rho$) at zero temperature for $U_2=0, U_0=12 zt$ (we set the
energy scale $zt=1$ throughout our calculation) for increasing
chemical potential along a vertical cross section in the ($\mu,U_0$)
phase diagram.  This trajectory traverses the superfluid regions with
non-integer $\rho$ as well as Mott plateaux with density fixed at
integral values $\rho=1$ and $2$. We observe that at $T=0$, although
the entropy (S) is zero in the superfluid (SF) region, it is nonzero
in the Mott insulator (MI) lobes. For $\rho=1$, the entropy is
$\ln(3)$, and for $\rho=2$ the entropy $S=\ln(6)$.  This nonzero
ground state entropy can be understood as follows. In the mean-field
treatment, the system Hamiltonian is a sum of single site Hamiltonians
and in the MI phase for one particle per site, three degenerate spin
components $\sigma=-1,0,1$, contribute to $S=\ln(3)$. For the $\rho=2$
Mott phase, the two spin-1 bosons on a single site have total spin
$F=0,2$, with $F=1$ eliminated by the symmetry constraint on the spin
functions.  The number of degenerate components is therefore $5$ due
to $F=2$, and $1$ due to the spin singlet $F=0$, so that $S=\ln(6)$.

The entropy for the ferromagnetic case $U_2<0$, $U_2=-0.1U_0,\,
U_0=12$ at $T=0$ is shown in Fig.~\ref{fig:entropyU2}(b).  A negative
value of the spin coupling, $U_2$, favors maximal total spin.  Hence,
$F=2$ for $\rho=2$, and $S=\ln(5)$ is reduced from its $U_0=0$
value. For $\rho=1$, {\it i.e.} the first Mott lobe, the entropy is
still $\ln(3)$ as before, and zero in the superfluid regime.
Antiferronetic values, $U_2>0$, favor the $F=0$ singlet phase for
$\rho=2$ and, therefore, $S=0$ for the second Mott
lobe. Fig.~\ref{fig:entropyU2}(c) shows the entropy for
$U_2=0.1U_0,\,U_0=12$ at $T=0$. As in the earlier cases, the entropy
is zero in the SF phase, and $\ln(3)$ in the first MI lobe. The
presence of finite entropy for the MI phase in mean-field theory even
at $T=0$, influences the topology of the isentropic curves, as we
shall discuss.

%%%%%%%%%%%%%%%
\begin{figure*}[ht]
\includegraphics[width=0.9\textwidth,angle=0]{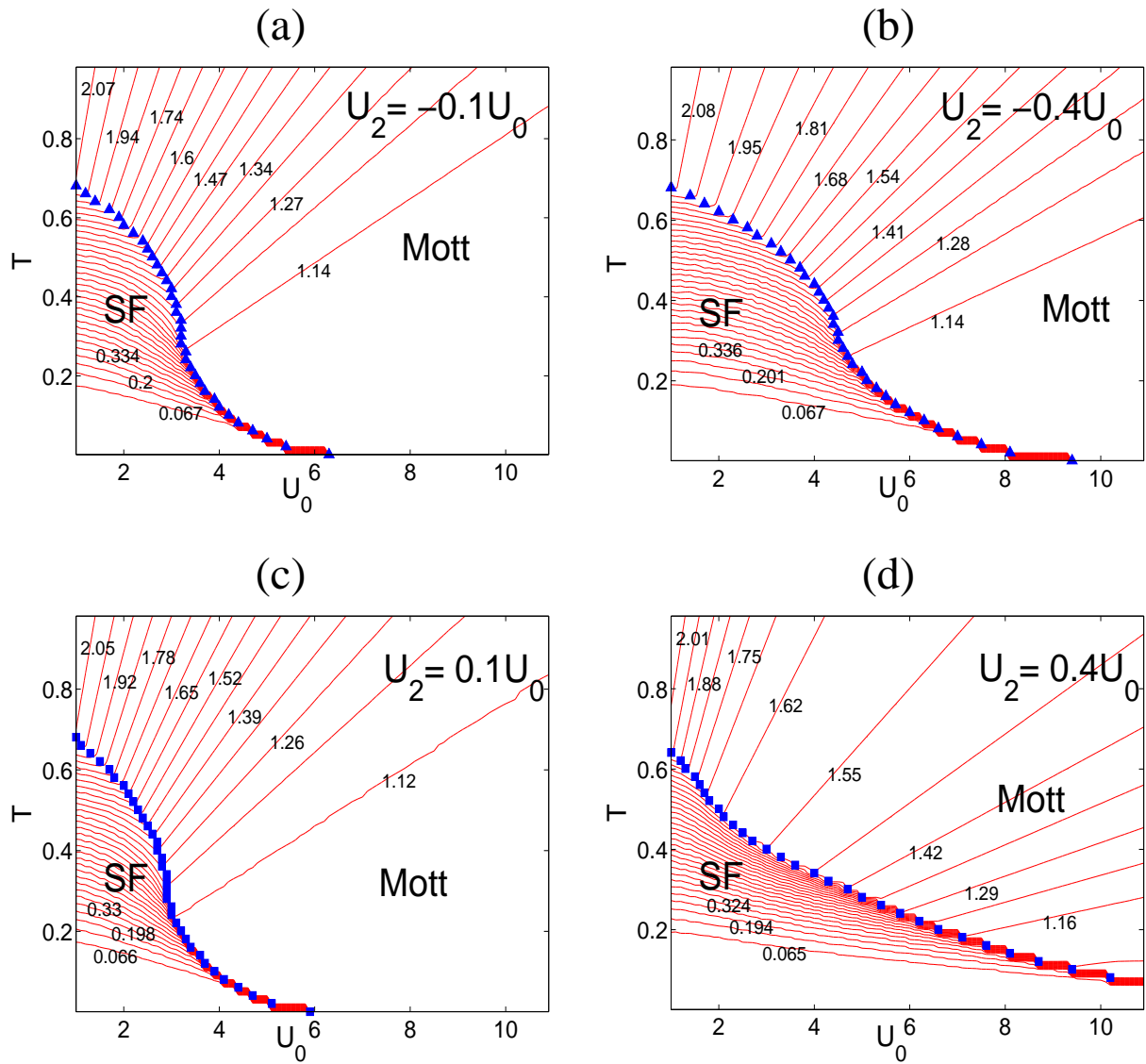}
\caption{\label{fig:isen2} (color online).  Isentropes overlaid on the
finite temperature phase diagram for ferromagnetic ($U_2<0$) and
antiferromagnetic ($U_2>0$) interactions.  The chemical potential
$\mu$ is chosen so that density $\rho=1$. For small values of $U_2$,
(a) $U_2=-0.1U_0$ and (c) $U_2=0.1U_0$, the isentropes are not much
different from each other and from $U_2=0$ (Fig.~3). For larger values
of $U_2$, (b) $U_2=-0.4U_0$ and (d) $U_2=0.4U_0$, the difference is
more visible. Along an isentropic line, first we have cooling followed
by heating for large $U_0$. The cooling and heating sepatrix again
lies on the superfluid-Mott phase boundary.}
\end{figure*}
%%%%%%%%%%%%%%%%

In contrast to spin-0 bosons, spin-1 bosons in a lattice exhibit both
first and second order phase transitions at finite temperature.  Here
we examine the entropy across these different phase boundaries. In
Fig.~\ref{fig:order}(a), discontinuities in entropy, density, and
superfluid density are evident across a first order phase boundary in
the ($\mu,T$) plane at a constant $T=0.1$, and $U_0=12, U_2=0$.  Such
first order finite temperature phase transitions should follow the
Clausius-Clapeyron equation from thermodynamics which relates the
discontinuity in the entropy and other parameters to the slope of the
phase boundary.  We have verifed that our results indeed satisfy this
identity.  For example, in Fig.~\ref{fig:order}(a), we find $\Delta
S/\Delta \rho = 3.10$ which equals the slope $d\mu/dT$ in the
($\mu,T$) plane.

As we increase the temperature, the entropy jump decreases, and
finally vanishes at the tricritical point. The entropy is continous
afterwards across a second order phase transition as we show in
Fig.\ref{fig:order}(b) for $T=0.2$.

To investigate further the effect of adiabatic ramping of the
optical lattice, we calculate the entropy at different couplings
(fixed $U_2$ and many $U_0$) and temperatures, and construct the
isentropic curves in the ($T,U_0$) plane. Fig.~\ref{fig:isen1}(a)
shows the isentropic curves for $U_2=0$ for fixed occupation
number $\rho=1$. Isentropic curves are overlaid on the finite
temperature phase diagram where the boundary separates $\rho_s>0$
(SF phase) and $\rho_s=0$ (MI/Normal phase). During adiabatic
ramping of the lattice, the system heats or cools following one of
these isentropic lines. For entropy $1.13$ at $U_0=0$ the
temperature starts at $T=0.43$ but decreases along the isentrope
as $U_0$ rises. At the SF boundary, $T$ begins to rise again.  In
fact, all the mean-field isentropes that we obtain show this
pattern - there is cooling in the SF regime and heating in the MI
regime with the heating-cooling separatrix exactly on the phase
boundary.

The occurence of cooling in the SF phase and heating in the Mott
phase, and the location of the heating-cooling separatrix exactly
on the phase boundary can be understood with the following
physical argument - moving away from the phase boundary towards
higher $U_0$, one enters the Mott phase with reduced number
fluctuations and an integer number of atoms per site. So, as $U_0$
increases at constant $T$, the entropy is reduced. Therefore, if
we want to keep the entropy constant as $U_0$ increases, the
temperature must rise, and there is heating in the Mott phase.
Similarly, moving away from the phase boundary towards the SF
regime by decreasing $U_0$, more and more particles enter the
condensate,  reducing the quantum depletion. More particles
entering the condensate means entropy is decreasing since the
condensate carries no entropy. So, to keep S constant as $U_0$
decreases in the SF, there must again be heating in the system.

The fact that within mean field theory phase boundaries also demark
the switch between cooling and heating isentropes is a phenomenon that
has also been observed in studies of classical models of nuclear
magnetism \cite{mouritsen88}.  Recently, it has also been shown that
the spin-one Blume-Capel model exhibits this same
behavior\cite{cone09}.  In both cases, it is found that in exact Monte
Carlo calculations for the same models the heating/cooling separatrix
continues to track the phase boundary qualitatively, but increasingly
breaks away as the temperature increases.

Another feature of the isentropic lines is that for low initial
entropy (which is also at low temperature) they terminate at the SF-MI
phase boundary without ever entering the MI phase.
Fig.~\ref{fig:entropyU2}, showed that within site-decoupled mean-field
theory, the entropy in the Mott regime with $\rho=1$ is
$\ln(3)=1.0986$ even at $T=0$.  As a consequence, if we follow an
isentrope with initial entropy less than $\ln(3)$ in the SF phase, as
$U_0$ increases beyond the critical value, the isentrope cannot reach
the MI phase because of its high ground state spin entropy.

%%%%%%%%%%%%%%  FIGURE  %%%%%%%%%%%%%%%%%%%%%%%%%%%%%%%%
\begin{figure}[ht]
\includegraphics[width=0.5\textwidth,angle=0]{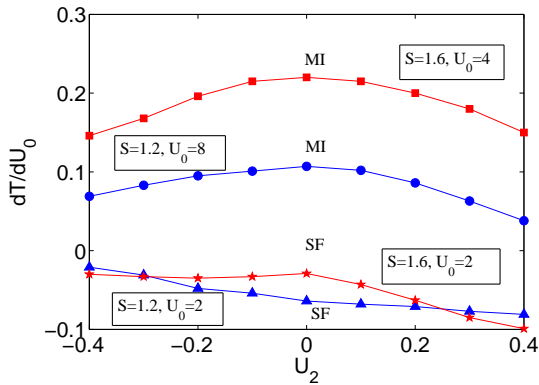}
\caption{\label{fig:u2} (color online).  The effect of spin coupling
$U_2$ on the rate of temperature change. As we vary $U_2$, the slopes
of the isentropes in the ($T,U_0$) plane change.  Here we show the
rates of change of temperature with $U_0$ for two isentropes in the
first Mott lobe and superfluid regime. In the MI region, the rate decreases
as we move away from $U_2=0$; while in the SF, it increases or decreases
depending on the value of $U_0$}
\end{figure}
%%%%%%%%%%%%%%%%%%%%%%%%%%%%%%%%%%%%%%%%%%%%%%%%%%%%%%%%

Fig.~\ref{fig:isen1}(b) shows entropy as a function of $U_0$ at constant
temperature intervals. As $U_0$ increases into the MI at constant
temperature, the low entropy values go to $S \geq 1.0986$. This illustrates why
only the isentropes with $S>1.0986$ in Fig.~\ref{fig:isen1}(a) continue to the
MI phase. This
termination of the isentropic lines is a shortcoming of the mean-field
theory treatment. There has not been any exact numerical study of isentropic
lines for the spin-1 model within QMC or DMRG.

In the Mott phase with independent sites, any of the three choices
$\sigma=0,\pm 1$ is possible on each site.  However, it is clear that
this degeneracy is broken in perturbation theory in $t$.  Two adjacent
sites with the same $\sigma$ will have a second order lowering $\Delta
E^{(2)}=-8 t^2/U$ while adjacent sites with different $\sigma$ will
have a second order lowering $\Delta E^{(2)}=-4 t^2/U$.  The entropy
per site is lower than $\ln(3)$.  Thus while mean-field theory
successfully captures the zero temperature phase diagram as has been
verified with QMC~\cite{batrouni09}, and predicts many qualitatively
correct results for the finite temperature phase diagram, it cannot
capture the low temperature entropy curves in the Mott region.

Next we examine the isentropes at a non-commensurate filling.
Fig.~\ref{fig:noninteger} shows the isentropic curves for $\rho=0.7$
overlaid on the finite temperature phase diagram. As in the
commensurate filling case, here we also find that the heating-cooling
separatrix is on the SF-Normal phase boundary. However, the low
temperature isentropes here do not terminate at the phase
boundary. Since the phase boundary for noninteger fillings never
touches the $T=0$ line, the isentropes can remain under the phase
boundary for arbitrary $U_0$.

We present isentropes for $U_2 \neq 0$ at commensurate filling
$\rho=1$ in Fig.~\ref{fig:isen2} which shows the isentropic curves for
both ferromagnetic and antiferromagnetic cases for two different spin
couplings.  We also show the SF-MI phase boundaries on the same plot.
The central observations are that the isentropes here have the same
general property as for $U_2=0$ - there is cooling in the SF phase and
heating in the MI phase, the phase boundary coincides with the
heating-cooling separatrix, and the low entropy lines do not enter the
Mott phase. For $U_2=-0.1U_0$ and $U_2=0.1U_0$, the isentropes are not
much different from each other and from the $U_2=0$ case. However for
larger magnetic coupling, $U_2=-0.4 U_0$ (Fig~\ref{fig:isen2}.(b)), we
find that adiabatic ramping results in a slower rate of cooling and
heating. Starting with the same initial temperature, and ending up at
the same final optical lattice depth in the MI phase, the temperature
is lower for $U_2=-0.4U_0$ than for $U_2=-0.1U_0$. This is also true
for the polar phase $U_2=0.4U_0$ in Fig~\ref{fig:isen2}(d).

%%%%%%%%%%%%%%%%%%%%%%%%%%%%%%%%%%%%%%%%%%%%
\begin{figure}[h]
\begin{center}
  \includegraphics[width=0.4\textwidth,angle=0]{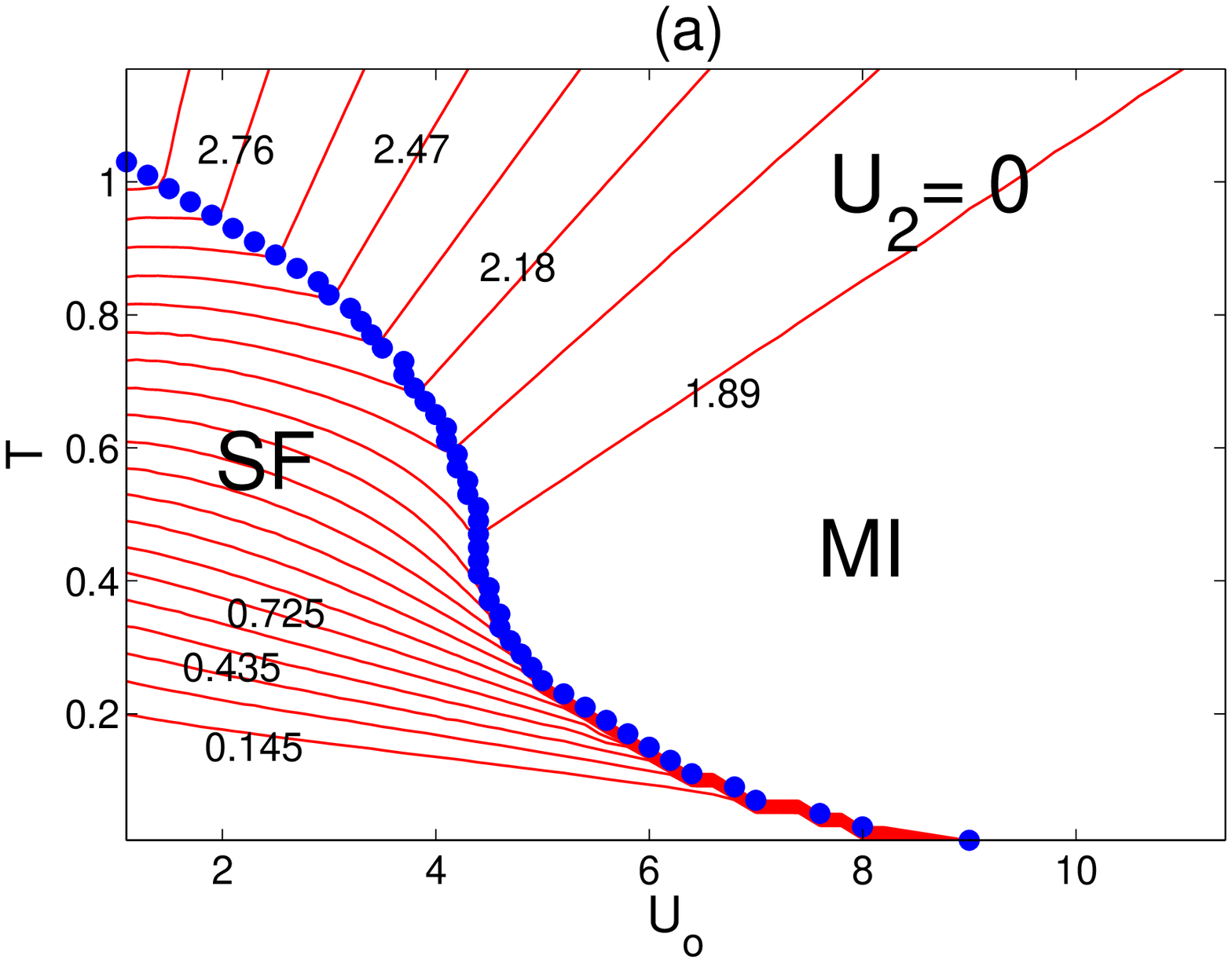}
\end{center}
\begin{center}
  \includegraphics[width=0.4\textwidth,angle=0]{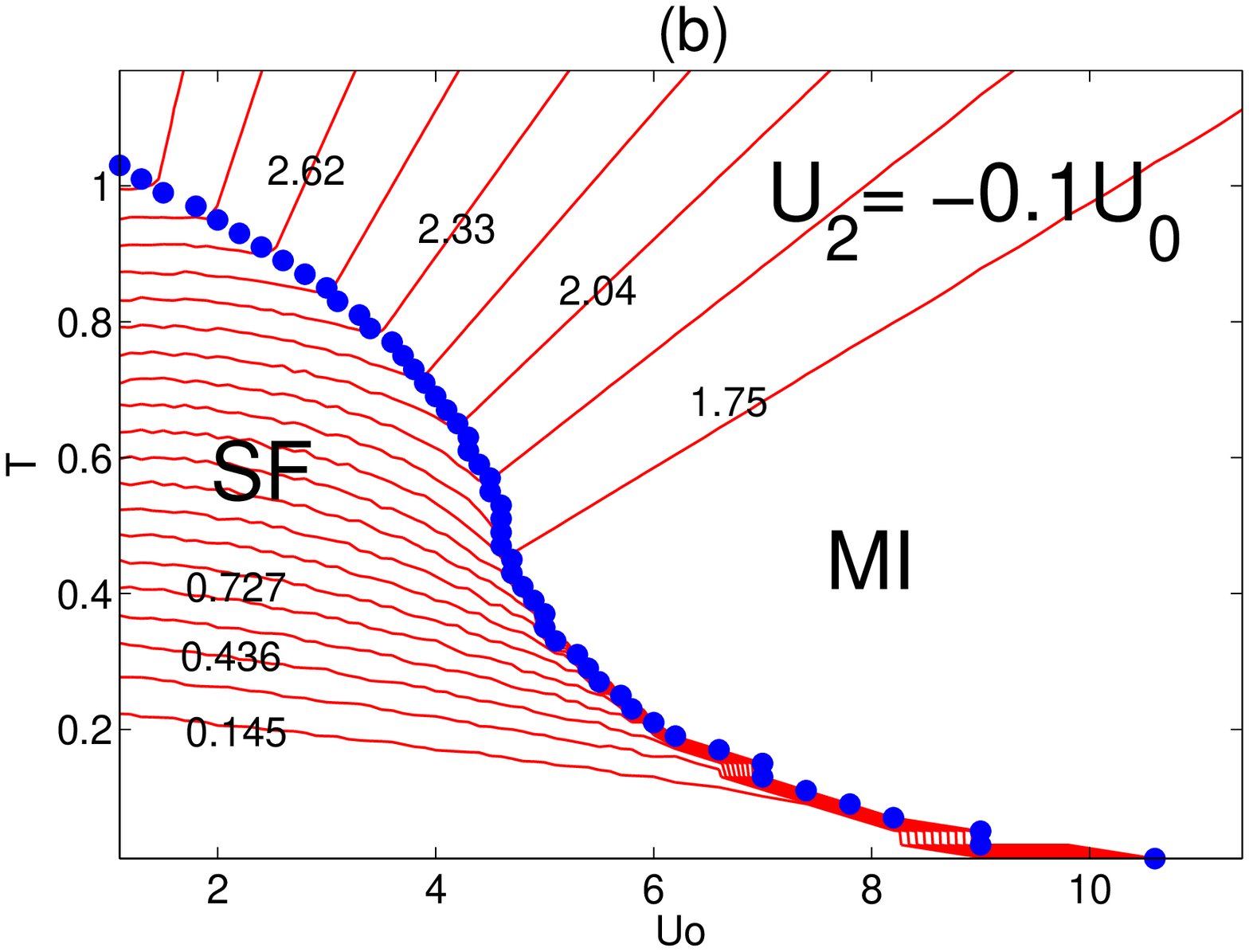}
\end{center}
\begin{center}
  \includegraphics[width=0.4\textwidth,angle=0]{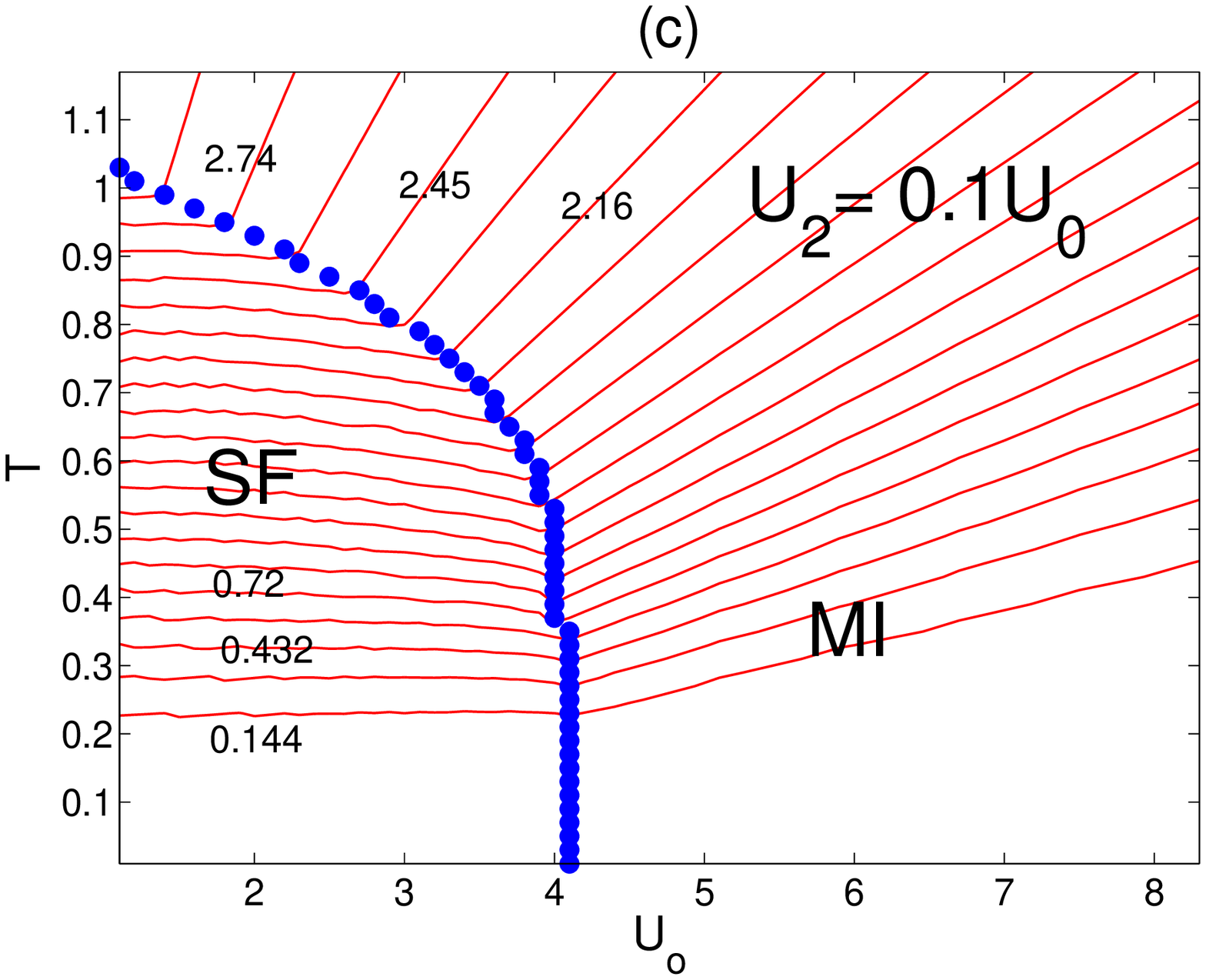}
\end{center}
\vspace{-0.6cm}
\caption{\label{fig:isen3} (color online).  Isentropes overlaid on the
finite temperature phase diagram for ferromagnetic ($U_2<0$) and
antiferromagnetic ($U_2>0$) interactions for density $\rho=2$. The
same values of $U_2$, $U_2=-0.1U_0$ and $U_2=0.1U_0$ were used as in
the $\rho=1$ results in Fig.~4.  There are several differences from
the case of unit filling.  Here, the sign of $dT/dU_0$ can change
within the superfluid.  The isentropes no longer exhibit pure
cooling. In addition, the antiferromagnetic isentropes in (c) have no
terminations on the phase boundary.}
\end{figure}
%%%%%%%%%%%%%%%%%%%%%%%%%%%%%%%%%%%%%%%%%%%%%

%%%%%%%%%%%%%%  FIGURE  %%%%%%%%%%%%%%%%%%%%%%%%%%%%%%%%
\begin{figure}[ht]
\includegraphics[width=0.5\textwidth,angle=0]{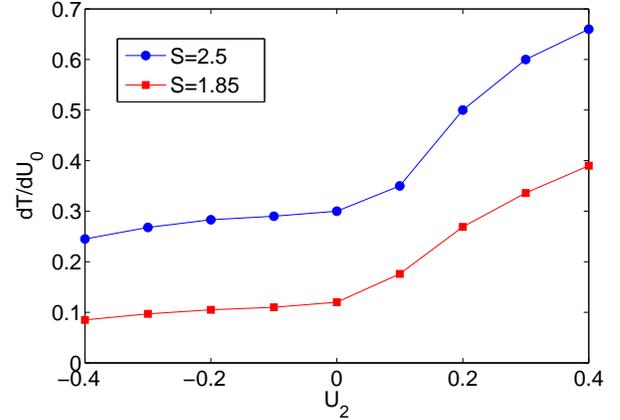}
\caption{\label{fig:u2rho2} (color online). The effect of spin
coupling $U_2$ on the rate of temperature change in the second
Mott lobe ($\rho=2$). Here we show the heating rate for two
isentropes S=2.5 and S=1.85. Unlike the $\rho=1$ case, the heating
rate is not maximum at $U_2=0$, and increases for $U_2>0$.}
\end{figure}
%%%%%%%%%%%%%%%%%%%%%%%%%%%%%%%%%%%%%%%%%%%%%%%%%%%%%%%%

In Fig.~\ref{fig:u2}, we show the slope of the isentropes
$dT/dU_0$, {\it i.e.} the heating rate in the Mott and SF phases,
as a function of $U_2$. We show results for two isentropes,
$S=1.6$ and $S=1.2$ in the MI and SF regime. We see that the
heating rate in the MI regime continually decreases as the
magnitude of $U_2$ increases.  In the SF region, the rate is
negative denoting cooling. For the two isentropes we show, the
characteristics of the rates differ based on whether we are in a
low or higher entropy curve. For the species of atoms used in
current experiments, $^{23}$Na and $^{87}$Rb, $|U_2|$ is on the
order of $0.03$. Our mean-field study indicates that the effect of
magnetic couplings of this magnitude on the adiabatic heating and
cooling rates is small.

From Fig.~\ref{fig:u2}, we see that the rate of heating in the
first Mott lobe is maximum at $U_2=0$, and slowly decreases as
magnetic coupling is turned on for both ferromagnetic and
anti-ferromagnetic case. As noted earlier, increasing $U_0$ in the
Mott region reduces fluctuations, creating order and thereby
reducing entropy. Thus the system must heat as $U_0$ is increased
to keep the entropy constant. Now if $U_2 \neq 0$, $U_2$ will be
trying to establish another kind of order. For example, in the
first Mott lobe, $U_2>0$ tries to establish a bond order in one
dimension, and $U_2<0$, a ferromagnetic order. These other orders
compete with the simple Mott insulator, slow down the reduction of
fluctuations as $U_0$ is increased, and result in a slower heating
rate.

In the spinless Bose-Hubbard model, the SF-MI transition in the $\rho=1$
and $\rho=2$ Mott lobes are qualitatively similar as they are both a
second order transition. The presence of magnetic coupling in the spin-1
model changes this scenario~\cite{pai08}. For $U_2>0$, the SF-MI
transition in the $\rho=2$ Mott lobe becomes a first order transition.
Phase boundaries for the even and odd Mott lobes also change as a
function of $U_2$~\cite{pai08,batrouni09}.

Our final results address the isentropes for density $\rho=2$.  In
Fig.~\ref{fig:isen3} we choose $U_2=0$, $U_2=-0.1U_0$, and
$U_2=0.1U_0$. Here also we see cooling in the SF and heating in
the Mott regime. Turning on the ferromagnetic coupling in
Fig.~\ref{fig:isen3}(b) does not have a huge impact on the
isentropes. However, for antiferromagnetic coupling $U_2=0.1U_0$
in Fig.~\ref{fig:isen3}(c), we see that the isentropes are visibly
different as is the shape of the phase boundary. Inside the SF,
the system cools down very slowly. Fig.~\ref{fig:u2rho2} shows the
heating rate $dT/dU_0$ for a constant entropy (S=2.5) in the
second Mott lobe as $U_2$ is turned on.  Unlike the case for the
first Mott lobe (Fig.~\ref{fig:u2}), there is no maximum at
$U_2=0$, and the heating rate increases monotonically with  $U_2$.
For $U_2>0$ and $\rho=2$, the SF/Mott phase boundary is first
order, and a spin singlet phase is formed in the Mott region.  The
appearance of order in the singlet phase greatly affects the
heating properties.

\vskip0.05in
\section{Conclusion}

In this paper we have extended previous mean field theory treatments
of the spin-1 Bose Hubbard model to compute the isentropic lines.
These quantities are of considerable experimental importance to
understand thermometry in ultra-cold atomic gases in an optical
lattice.

We presented the isentropes in the temperature-interaction
strength ($T,U_0$) plane for ferromagnetic, antiferromagnetic, and
zero spin couplings. Following these isentropic lines, temperature
changes can be determined during adiabatic loading of spin-1
bosonic atoms in an optical lattice. The isentropes have a number
of interesting features. First, they exhibit pure cooling within
the SF and heating in the normal/Mott phase.  The phase boundary
precisely corresponds to the location of the change in sign of
$dT/dU_0$, i.e. the heating-cooling separatrix lies on the SF-MI
phase boundary. This can be understood in the symmetric view of
heating as one moves away in $U_0$ from the phase boundary in
either direction of SF/Mott. The system gets more ordered with
reduced entropy. And therefore, temperature must increase to keep
the entropy constant. Second, the low entropy (low temperature)
isentropes terminate on the phase boundary because of the nonzero
ground state entropy which mean field theory gives for the Mott
phase. For noncommensurate fillings, the isentropes show similar
characteristics except that the low entropy lines do not terminate
at the phase boundary.

The effect of ferromagnetic and antiferromagnetic couplings on isentropes
has been examined, and we quantify the rate of heating as $U_2$ changes.
We find that in the experimentally relevant regime of $|U_2|=0.03$ for $^{23}$Na
and $^{87}$Rb, the changes in heating-cooling rate is not very different
from the $U_2=0$ case. It would be interesting to extend QMC
simulations of the magnetic and superfluid properties on the spinor
Bose-Hubbard model to study these thermodynamic properties as well.

\begin{acknowledgments}
We thank Dave Cone and Rajiv Singh for fruitful discussions, and the referee for useful observations on our calculation. This work was supported under ARO Grant No. W911NF0710576 with funds from DARPA
OLE program. GGB is supported by the CNRS (France) PICS 3659.

\end{acknowledgments}

{}

\end{document}